\documentclass[hidelinks]{interact}

\usepackage{epstopdf}% To incorporate .eps illustrations using PDFLaTeX, etc.
\usepackage[caption=false]{subfig}% Support for small, `sub' figures and tables
\usepackage{upgreek}
\usepackage{multirow}
\usepackage{color}
\usepackage{ulem}
\usepackage{graphicx}
\usepackage{subfig}
\usepackage{tikz}
\usepackage{hyperref}
\usepackage[longnamesfirst,sort]{natbib}% Citation support using natbib.sty
\bibpunct[, ]{(}{)}{;}{a}{,}{,}% Citation support using natbib.sty
\theoremstyle{plain}% Theorem-like structures provided by amsthm.sty
\newtheorem{theorem}{Theorem}[section]
\newtheorem{assumption}{Assumption}

\newtheorem{proposition}[theorem]{Proposition}

\theoremstyle{definition}

\newtheorem{example}[theorem]{Example}

\theoremstyle{remark}

\def\logit{\text{logit}}

\def\pr{p}

\def\ind{\text{ }\begin{picture} (9,8)
         \put(0,0){\line(1,0){9}}
         \put(3,0){\line(0,1){8}}
         \put(6,0){\line(0,1){8}}
         \end{picture}\text{ }
        }

\def\nind{\text{ }\begin{picture}(9,8)
         \put(0,0){\line(1,0){9}}
         \put(3,0){\line(0,1){8}}
         \put(6,0){\line(0,1){8}}
         \put(1,0){{\it /}}
         \end{picture}\text{ }
    }

\begin{document}

\articletype{Original article}% Specify the article type or omit as appropriate

\title{A confounding bridge approach for double negative control inference on causal effects}

\date{}
\author{
\name{Wang Miao\textsuperscript{a}\thanks{CONTACT Wang Miao. Email: mwfy@pku.edu.cn}, Xu Shi\textsuperscript{b},  Yilin Li\textsuperscript{a},  Eric J.Tchetgen Tchetgen\textsuperscript{c}}
\affil{\textsuperscript{a}Department of Probability and Statistics, Peking University; \textsuperscript{b}Department of Biostatistics, University of Michigan; \textsuperscript{c}Department of Statistics and Data Science, University of Pennsylvania}
}

\maketitle

\begin{abstract}
Unmeasured confounding is a key challenge for causal inference. In this paper, we establish a   framework  for  unmeasured confounding adjustment with negative control variables.
A negative control outcome  is associated with the  confounder but not causally affected by the  exposure in view, 
and  a negative control exposure is  correlated with the primary exposure or the  confounder  but does not  causally affect the outcome of interest.
We   introduce  an outcome confounding bridge function that  depicts  the  relationship between the confounding effects on the primary outcome   and the   negative control outcome,
and we incorporate   a negative control exposure to identify  the bridge function and the average causal effect. 
We also consider the extension  to the positive control setting by allowing for  nonzero causal effect of the primary exposure on the  control outcome.
We illustrate our approach with simulations and apply it to a   study about the short-term effect of air pollution on mortality. 
Although a standard analysis shows a significant acute effect of  PM2.5 on mortality,  our analysis  indicates that this effect may be confounded,
and after double negative control adjustment, the effect  is  attenuated toward zero.

\end{abstract}


\section{Introduction}
Observational studies offer an important source of data for causal inference in socioeconomic, biomedical,  and epidemiological  research.
A major challenge for observational studies is the potential for confounding factors of the exposure-outcome relationship in view.
The impact of observed confounders on causal inference can be alleviated by direct adjustment methods such as inverse probability weighting, matching, regression, and doubly robust methods \citep{rubin1973use,rosenbaum1983central,stuart2010matching,bang2005doubly}.
However, unmeasured confounding is present in many observational studies.
In this case, causal effects cannot be uniquely determined by the  observed data without extra assumptions.
As a result, the aforementioned adjustment methods may be   biased and potentially misleading in the presence of unmeasured confounding.  
Sensitivity analysis methods \citep{cornfield1959,rosenbaum1983assessing}
are widely used  to evaluate the impact of  unmeasured confounding and to assess  robustness of causal inferences, 
but in general they cannot completely correct for confounding bias.
Auxiliary variables  are particularly useful to adjust for unmeasured confounding in observational studies.
The instrumental variable (IV)  approach \citep{wright1928tariff,goldberger1972,baker1994paired,robins1994correcting,angrist1996identification},
rests on  an auxiliary covariate that  (i)  has no direct effect on the outcome,  (ii) is independent of the unmeasured confounder, and (iii) is associated with the exposure. 
In addition,  a  structural  outcome  model or a monotone effect of the IV on the treatment,  is typically required to identify a causal effect. Although the IV approach has   gained popularity in causal inference literature in recent years,  particularly in health and social sciences, the approach is highly sensitive to  violation of any of assumptions (i)--(iii).

A more recent framework that leverages both negative control exposures and negative control outcomes to mitigate confounding bias is known as proximal (or negative control) inference \citep[e.g.,][]{miao2018proxy, tchetgen2020introduction}, where a negative control outcome is an outcome variable that is associated with the confounder but not causally affected by the primary exposure,
and a negative control exposure is an exposure variable that is correlated with the primary exposure or the confounder  but does not  causally affect the outcome of interest. 
Starting from an earlier version of this paper \citep{miao2018negative},  we develop an outcome  confounding bridge framework for identification and inference about causal effects by using a pair of negative control exposure and outcome to account for unmeasured confounding bias.
This line of work contributes to the literature by %relaxing  previous stringent model 
providing an alternative identification assumption, proposing practical inference methods, and establishing connections to conventional  approaches for confounding bias adjustment.
A subsequence  of related methods have been developed for categorical cases \citep{shi2020multiply}, dynamic treatment regime \citep{qi2022proximal}, heterogeneous treatment effect \citep{sverdrup2023proximal}, longitudinal studies \citep{ying2023proximal}, mediation analysis \citep{dukes2021proximal}, outcome-dependent sampling \citep{li2022doubly},  reinforcement learning \citep{bennett2021proximal}, semiparametric theory \citep{ cui2020semiparametric}, survival analysis \citep{ying2022proximal}, and synthetic control \citep{shi2021theory}.

In this paper,  we  illustrate the outcome confounding bridge framework for identification and inference about causal effects.
Our approach is based on a key assumption that the confounding effect on the primary outcome matches 
that on a transformation of the negative control outcome; throughout, 
this transformation is referred to as an outcome confounding bridge function,
which is formally introduced in  Section 3. 
Although in practice the  bridge function is unknown,  it can be   identified by using  a negative control exposure under certain completeness condition. Consistent and asymptotically normal estimation of the average causal effect can be  achieved by   the generalized method of moments described in Section 4. 
In Section 5,  we generalize the negative control approach by allowing for a positive control outcome, which can be causally affected by the primary exposure. We also develop a sensitivity analysis approach for checking the robustness of causal inference against the negative control assumptions.
In Section 6, we conduct  simulation studies to evaluate the performance of the double negative control approach and compare it to competing methods.
In Section 7, we apply our  approach to a time-series study about the effect of air pollution on mortality.
We conclude in Section 8 with discussion about implications of our  approach in observational studies and modern data science.

\section{Definition and examples of negative control outcomes}
Throughout, we let $X,Y,$ and $V$ denote  the primary  exposure,     outcome, and a vector of observed covariates, respectively. 
Vectors are assumed to be column vectors, unless explicitly transposed.
Following the convention in causal inference,  we use  $Y(x)$ to denote the potential outcome under  an intervention which sets $X$ to  $x$,  and maintain  consistency assumption that the observed outcome is  a realization of the    potential outcome under  the exposure actually received: $Y=Y(x)$ when $X=x$.  
We focus on the average causal effect (ACE) of $X$ on $Y$, which is  a contrast of the potential outcome mean between two exposure levels, for instance, ${\rm ACE}_{XY}=E\{Y(1)-Y(0)\}$ for a binary exposure.

The ignorability assumption   stating that $Y(x)\ind X\mid V$ is  conventionally made    in causal inference,  
but it does not hold  when unmeasured  confounding is present. In this case,     latent ignorability that states $Y(x)\ind X\mid (U,V)$ is more reasonable,  
allowing for  an unobserved confounder  $U$.
For notational convenience, we  present results conditionally on observed covariates and  suppress $V$  unless otherwise stated.
\begin{assumption}[Latent ignorability]\label{assump:cnfd}
$Y(x)\ind X\mid U$ for all $x$.
\end{assumption}
Given   latent ignorability, we have that for all  $x$,
\begin{eqnarray}\label{eq:pd}
E\{Y(x)\}=E\{ E(Y\mid U, X=x)\}. 
\end{eqnarray}
The crucial  difficulty of implementing  \eqref{eq:pd} is that $U$ is not observed and both the conditional mean $E(Y\mid U,X=x)$ and  the  density function  $\pr(U)$ are  not identified.

We introduce negative control variables to mitigate the problem of unmeasured confounding. 
Suppose   an auxiliary  outcome $W$ is available and satisfies  the following assumption.
\begin{assumption}[Negative control outcome]\label{assump:nco} $W\ind X \mid U$ and $W\nind U$.
\end{assumption}
The assumption realizes the notion of  a negative control outcome that it is  associated with the confounder but  not causally  affected by the primary exposure. 
Moreover, the sets of unmeasured confounders for $(X,W)$ and $(X,Y)$ are the same, which corresponds to the U-comparable assumption of \cite{lipsitch2010negative}. 
Assumption \ref{assump:nco}  does not impose restrictions on  the association of $W$--$Y$.
A special case  is the  nondifferential assumption of \cite{lipsitch2010negative} and \cite{tchetgen2014control}, which further requires $W\ind Y\mid U$ and does not allow for extra confounders of  $W$--$Y$  association. 
Justification of  Assumption  \ref{assump:nco} and choice of negative controls  require subject matter knowledge.

\begin{example}\label{exmp:1}
In a study about the effect of  acute stress  on  mortality from heart disease,  \cite{trichopoulos1983psychological}  found increasing  mortality from cardiac and external causes during the days immediately after the 1981 earthquake in Athens. However, acute stress due to the earthquake is unlikely to quickly cause  deaths from cancer.
In a parallel analysis, they found no increase in  risk of  cancer mortality, which is evidence in favor of 
no confounding and  reinforces their claim that acute stress  increases  mortality from heart diseases. In this study, the exposure is the psychological stress post-earthquake, and the outcome of interest is deaths from cardiac events. The unmeasured confounder can be the nutritional level or economic status which influences both the exposure and outcome. However, the exposure does not directly influence other causes of death such as cancer. The cancer mortality is the  negative control outcome, and   they are used to test whether confounding bias is present and  to  evaluate the plausibility of a causal association. 
\end{example}

However, it is far more challenging to identify a causal effect  with  a single negative control outcome. In the Supplementary Material, we provide two distinct parameter values of a fully parametric model that lead to identical  distribution of $(X,Y,W)$.
In the next section, we explore more realistic conditions under which identification can be achieved.

\section{Identification of causal effects with a negative control pair}

%In Example \ref{exmp:ct}, although $\beta$ cannot be identified  solely by the distribution of $(X,Y,W)$,  we observe that once  the ratio $\alpha_3/\alpha_1$ is known,  $\beta$ is identified by $\beta = \partial E(Y\mid X=x)/\partial x- \alpha_3/\alpha_1\times \partial E(W\mid X=x)/\partial x$. 
%The fact that $(\alpha_1,\alpha_3)$ encode the  confounding effects of $U$ on $W$ and $Y$, respectively,  motivates us to introduce the  outcome confounding bridge function.

The literature of proximal causal inference indicates  it is possible to identify the average treatment effect with both a negative control outcome and a negative control exposure, 
by invoking  an outcome confounding bridge function  \citep{miao2018proxy,cui2020semiparametric} to characterize relationship between the confounding effects on both the primary outcome and the negative control outcome.

\begin{assumption}[Outcome confounding bridge]\label{assump:bridge}
There exists some  function $b(W,X)$ such that for all $x$,
\begin{eqnarray}\label{eq:bridge}
E (Y\mid U,X=x) =E\{ b(W,x)\mid U,X=x\}.
\end{eqnarray}
\end{assumption}
When covariates $V$ are observed,  \eqref{eq:bridge} becomes $E \{Y\mid U,V,X=x\}  =E\{ b(W,V,x)\mid U,V,X=x\}$.
Assumption \ref{assump:bridge}  states that the confounding effect of $U$ on $Y$ at  exposure level $x$, is equal to the  confounding effect  of $U$ on the variable $b(W,x)$, a transformation of $W$; it goes beyond   U--comparability by characterizing  the relationship between the confounding effects of $U$ on $Y$ and  $W$. 
%As a special case, when  $U$ is binary, the assumption holds as long as $W\nind U$.
We  illustrate the assumption  with an example of the linear outcome confounding bridge.
Assuming that  $E(Y\mid U,X)=(1, X, U, XU)\beta$ and that $E(W\mid U)$ is linear in  $U$, then \eqref{eq:bridge} holds with 
$b(W,X;\gamma)= (1, X, W, XW)\gamma$, for an appropriate value of $\gamma$.
Linearity in $W$ in this bridge function, corresponds to  a proportional relationship between  the confounding effects of $U$ on $Y$ and $W$.
\iffalse
If  interaction does not occur, then  the outcome confounding bridge reduces to an additive form, as in  Example \ref{exmp:ct}.

\begin{example}[Additive and multiplicative outcome confounding bridge]\label{exmp:add}
For an additive data generating process, $E(Y\mid U,X)=b_1(X)+U$,  \eqref{eq:bridge} holds with  an additive bridge function,
$b(W,X)=b_1(X)+b_2(W)$ if  $E\{b_2(W)\mid U\}=U$.
Analogously,  for a multiplicative data generating process, $E(Y\mid U,X)=\exp\{b_1(X)+U\}$,  \eqref{eq:bridge} holds with
$b(W,X)=\exp\{b_1(X)+b_2(W)\}$ if $E\{\exp(b_2(W))\mid U\}=\exp(U)$.
\end{example}

The  additive and  multiplicative data generating processes are  often assumed in empirical studies, with $b_1(x)$ encoding the causal effect on the mean  and the risk ratio scales, respectively. These examples  demonstrate the relationship between the data generating process and the outcome confounding bridge.  
\fi

The average causal effect can be recovered by integrating the outcome confounding bridge over $W$.  

\begin{proposition}\label{prp:1}
Given Assumptions  \ref{assump:cnfd}--\ref{assump:bridge},   we have  that for all $x$,
\begin{eqnarray}\label{eq:idn}
E\{Y(x)\} & = &  E\{b(W,x)\}.
\end{eqnarray}
\end{proposition}
The proposition   reveals the   role of the  negative control outcome and the outcome confounding bridge $b(w,x)$.
Given $b(w,x)$, the potential outcome mean and the average causal effect can be  identified without an additional assumption.
We emphasize that without knowledge of such bridge function, identification is not possible in general, even under a fully parametric model  and  full knowledge of  the confounder distribution. However, in practice,  the outcome confounding bridge is unknown.
In order to identify the confounding bridge, we introduce an auxiliary  exposure variable named negative control exposure $Z$ that satisfies  the following exclusion restrictions.
\begin{assumption}[Negative control exposure]\label{assump:nce}
$Z\ind Y\mid (U,X)$ and $Z\ind W\mid (U,X)$.
\end{assumption}
The assumption  states that  upon  conditioning on  the primary exposure and the confounder, $Z$  does not affect either the primary outcome $Y$  nor the negative control outcome $W$.
This assumption does not impose  restrictions  on  the association between $Z$ and $X$ and allows $Z$ to be confounded.
A special case  is the instrumental variable \citep{wright1928tariff,goldberger1972} that is independent of 
the confounder,  in addition to the exclusion restrictions. 
Below we provide an empirical example for  negative control exposures.

%Researchers have considerable interest in  the effects of intrauterine exposures on   offspring outcomes, for example, the effects of maternal smoking, distress, and diabetes  during pregnancy on offspring birthweight, asthma, and    adiposity. If there are causal intrauterine mechanisms, then maternal exposures are expected to  have  an  influence on offspring outcomes, but  conditional on maternal exposures, paternal exposures should not affect offspring outcomes.  Thus, paternal exposures are used as negative control exposures.
%For instance, \cite{dsmith2008assessing,dsmith2012}   used paternal  smoking as a negative control exposure to adjust  the   intrauterine influence of maternal smoking on offspring  birthweight and later-life body mass index. 
%

\begin{example}\label{exmp:air}
In a time-series study  about  air pollution, \cite{flanders2017new} used  air pollution level in future  days as  negative control exposures to test  and  reduce confounding bias.  For day $i$, let $X_i,Y_i,U_i$ denote the   air pollution level (e.g., PM2.5), a public health outcome (e.g., mortality), and the unmeasured confounder, respectively; although $Y_i$ is possibly  affected by   air pollution in the current and past days,  it is not affected by future days air pollution, $X_{i+1}$ for instance; moreover,  public health outcomes in general do not affect  air pollution in the immediate future. Thus,  it is reasonable to use  $X_{i+1}$   as  a negative control exposure.
\end{example}

Just as negative control outcomes, a negative control exposure can also be used to test whether  confounding bias  occurs by checking if $Z$ is independent of  $Y$ or $W$ after conditioning on $X$.
Alternatively, we propose to use a negative control exposure  to identify the outcome confounding bridge.  
Taking  expectation of $U$ with respect to $\pr(U\mid Z,X)$ on both sides of $E(Y\mid U,X)=E\{b(W,X)\mid U,X\}$, we obtain 
\begin{eqnarray}\label{eq:intg}
E(Y\mid Z,X)=E\{b(W,X)\mid Z,X\}.
\end{eqnarray}
The equation suggests that the outcome confounding bridge  also captures the relationship between the  crude effects  of $Z$ on $Y$ and $W$.
This is because conditional on $X$, the crude effects of  $Z$ on $(Y,W)$  are completely driven by the association with  the confounder $U$.
Equation \eqref{eq:intg}  offers a feasible strategy to identify the outcome confounding bridge   with a negative control exposure.  
Because $E(Y\mid Z,X)$ and $\pr(W\mid Z,X)$ can be obtained  from the observed data,  one can solve the equation  for the bridge function. This type of integral equation is known as the Fredholm integral equation of the first kind. Consider the case where both $W$ and $Z$ are binary, and then \eqref{eq:intg} becomes two linear equations with two unknown parameters.
The following  condition  concerning  completeness of $\pr(W\mid Z,X)$ guarantees uniqueness of the solution. 
\begin{assumption}[Completeness of $\pr(W\mid Z,X)$]\label{cndt:cmp}
For all $x$, $W\nind Z\mid X=x$; and for any square integrable function $g$, if $E\{g(W)\mid Z=z,X=x\}=0$ for almost all $z$, then $g(W)=0$ almost surely.
\end{assumption}
Completeness  is  a commonly-made  assumption in identification problems, such as  instrumental  variable identification discussed by \cite{newey2003instrumental}, \cite{d2011completeness}, \cite{darolles2011nonparametric}, and \cite{andrews2017examples}.
These previous results about completeness can  equally be applied here.
For a binary confounder, completeness holds as long as $W\nind Z\mid X=x$ for all $x$;
completeness also holds for many widely-used distributions such as   exponential families \citep{newey2003instrumental} and  location-scale families \citep{hu2018nonparametric}. However, if ACE is of primary interest, the uniqueness assumption is not a prerequisite for estimation and inference, as indicated in \cite{zhang2023proximal}.

\begin{theorem}\label{thm:idn2}
Under Assumptions \ref{assump:cnfd}--\ref{cndt:cmp}, Equation \eqref{eq:intg} has a unique solution,  and the potential outcome mean is identified  by plugging such solution into Equation \eqref{eq:idn}.
\end{theorem}

So far, under the completeness condition, we have identified the potential outcome mean  without imposing any model restriction on the outcome confounding bridge. 
If the  bridge function belongs to a parametric or semiparametric model, the completeness condition can be weakened.

\begin{theorem}\label{thm:idn3}
Under Assumptions \ref{assump:cnfd}--\ref{assump:nce} and given a  model  $b(W,X;\gamma)$ for the bridge function indexed by a finite or infinite dimensional parameter $\gamma$, if  for all $x$, $E\{b(W,x;\gamma) - b(W,x;\gamma')\mid Z,X=x\}\neq0$ with a positive probability for any    $\gamma\neq  \gamma'$, then  $\gamma$ is identified by solving $E\{Y - b(W,X;\gamma)\mid Z,X\}=0$, and thus   the potential outcome mean is identified.
\end{theorem}

For instance, the linear  model  $b(W,X;\gamma)=(1,X,W,XW)\gamma$  is identified as long as $E(W\mid Z,X)\neq E(W\mid X)$ with a positive probability, i.e., $W$ is not mean independent of $Z$ after conditioning on  $X$. 
Under the  linear outcome confounding bridge, the relationship between the  causal effect, the confounding bias, and  crude effects 
has an explicit form, as shown in the following example.

\begin{example}\label{exmp:neg2}
Consider  binary exposures $(X,Z)$ and the linear  confounding  bridge function, $b(W,X;\gamma)=\gamma_0+\gamma_1X +\gamma_2 W+\gamma_3XW$,  
and let  ${\rm RD}_{XY\mid Z}=E(Y\mid X=1,Z)-E(Y\mid X=0,Z)$    denote  the      risk difference  of $X$ on $Y$ conditional on $Z$;
then  $(\gamma_2,\gamma_3)$ are identified by
\begin{eqnarray*}
\gamma_2= \frac{{\rm RD}_{ZY\mid X=0}}{{\rm RD}_{ZW\mid X=0} },&& \gamma_3= \frac{{\rm RD}_{ZY\mid X=1}}{{\rm RD}_{ZW\mid X=1} } - \gamma_2.
\end{eqnarray*}
The average causal effect of $X$ on $Y$ is identified by
\begin{eqnarray*}
{\rm ACE}_{XY} &=& E({\rm RD}_{XY\mid Z})  - (\gamma_2+\gamma_3)E({\rm RD}_{XW\mid Z})\\ 
&& + \gamma_3 \sum_{z=0}^{1} \{{\rm RD}_{XW\mid Z=z} \times  \pr(Z=z, X=1)\}.
\end{eqnarray*}
If  the   bridge function is additive, i.e., assuming that $\gamma_3=0$, then $\gamma_2= E({\rm RD}_{ZY\mid X})/E({\rm RD}_{ZW\mid X})$ and 
\begin{eqnarray}\label{eq:ace}
{\rm ACE}_{XY}=E({\rm RD}_{XY\mid Z}) - \frac{E({\rm RD}_{ZY\mid X})}{E({\rm RD}_{ZW\mid X})} \times E({\rm RD}_{XW\mid Z}).
\end{eqnarray}

%{\red delta method to derive the variance using summary data, equal to 2SLS when $E(y\mid z,x),E(w\mid z,x)$ include no interaction. parameter of V  not identified}

\end{example}
This example  offers a convenient adjustment  when only summary data about  crude effects are available.
In the Supplementary Material, we extend this example  by allowing for   exposures of arbitrary type and   a nonparametric outcome confounding bridge. 
Identification of causal effect is also possible without completeness condition, see \cite{zhang2023proximal}.
%In the next section, we consider  estimation and inference methods when  individual-level  data are available.

So far, we have identified the average causal effect with  a pair of negative control exposure and outcome. 
If the treatment effect on the treated, $E\{Y(1)-Y(0)\mid X=1\}$, is of interest instead,  one only needs a weakened  outcome confounding bridge assumption imposed on  the control group, i.e., $E(Y\mid U,X=0)=E\{b(W)\mid U,X=0\}$ for some function $b(W)$, and then  a negative control exposure can be used to identify $b(W)$.
Our confounding bridge approach clarifies the roles of negative control exposure and outcome in confounding bias adjustment.
A negative control outcome is used to mimic  unobserved potential outcomes via the outcome confounding bridge that  captures the relationship between the effects of confounding. The confounding bridge approach unifies previous  bias adjustment methods  in the  negative control  design.
The approaches of \cite{tchetgen2014control} and \cite{sofer2016}  are special cases of our outcome confounding bridge approach by assuming   rank preservation of individual potential outcomes or monotonicity about the confounding effects. 
The factor analysis approach of  \cite{gagnon2013removing} and \cite{wang2017confounder}    in fact identifies the outcome confounding bridge via   factor loadings on the confounder. 
Therefore, these previous approaches reinforce the key role of the confounding bridge in   the  negative control  design.
Confounder  proxies used by \cite{miao2018proxy} and \cite{kuroki2014measurement} can be  viewed as special negative controls in our framework.
The identification strategy of \cite{miao2018proxy} rests on a completeness condition involving the unmeasured confounder; however, our completeness condition depends only on observed variables.
Our identification strategy rests on the outcome confounding bridge; 
alternatively, \cite{cui2020semiparametric} propose an  identification approach which rests on an exposure confounding bridge $e(Z,X)$ defined by the solution to $E\{e(Z, X) \mid W,  X=x\}=\{p(X=x \mid W)\}^{-1}$, connecting the negative control exposure to the inverse propensity score. Their identification is guaranteed by a completeness condition of $p(U\mid W,X)$.

\section{Estimation}
We focus on estimation of the average causal effect  $\Delta=E\{Y(x_1)-Y(x_0)\}$ that contrasts potential outcomes mean under  two exposure levels $x_1$ and $x_0$.
We first consider estimation with  i.i.d. data samples  and then generalize to time-series data.
Suppose that one has specified a parametric model  for the outcome confounding bridge, $b(W,V,X;\gamma)$. Practically, we recommend users start with a linear additive $b(W, X) = \gamma_1W + \gamma_2 X$ or exponential multiplicative $b(W, X) = \exp(\gamma_1W + \gamma_2 X)$. 
However, a misspecified low dimensional model $b(W, X)$   can potentially lead to a biased result.  
The users can use a variety of more flexible approaches such as semiparametric (e.g. partially linear model, single index model) or nonparametric (e.g. generalized additive, reproducing kernels, neural networks, see \citet{cui2020semiparametric, kallus2021causal}), to check the robustness of the estimated causal effect on $b(W,X)$, thus further alleviating concerns about misspecification bias. 
A standard approach to estimate  $\theta=(\gamma,\Delta)$ is     the generalized method of moments \citep{hansen1982large,hall2005generalized}.
We let $D_i=(X_i,Z_i,Y_i,W_i,V_i),1\leq i \leq n$ denote the observed data samples.  
Define the vector of  moment restrictions 
\begin{eqnarray}\label{eq:gmm}
h(D_i; \theta)= \left\{\begin{array}{l}
\{Y_i - b(W_i,V_i,X_i; \gamma)\}\times q(X_i,V_i,Z_i), \\
\Delta -\{b(W_i,V_i,x_1;\gamma)-b(W_i,V_i, x_0;\gamma)\},
\end{array}\right.
\end{eqnarray}
with a user-specified vector function $q$, and let $m_n(\theta)=  1/n\sum_{i=1}^{n} h(D_i;\theta) $; the   GMM  solves 
$
\widehat \theta = \operatorname{arg}\min_{\theta} m_n^\top(\theta)\ \Omega\  m_n(\theta),
$
with  a user-specified  positive-definite weight matrix $\Omega$. 
%The first component in \eqref{eq:gmm} consists of unbiased  estimating equations for $\gamma$ because
%$E\{Y-b(W,V,X;\gamma)\mid V,X,Z\}=0$, and the second one for $\Delta$ because $E\{Y(x)\}=E\{b(W,V,x;\gamma)\}$.
%For a bridge function having an additive form $b(W,V,X;\gamma)= b_1(X;\gamma_1) +b_2(W,V;\gamma_2)$ or a multiplicative one $b(W,V,X;\gamma)= \exp\{b_1(X;\gamma_1) +b_2(W,V;\gamma_2)\}$, 
%where  the structural parameter  $\gamma_1$ is  of interest,  only  the first component  of \eqref{eq:gmm} needs to be included when implementing the GMM. 
%Sometimes, it is of interest  to consider a semiparametric bridge function with a parametric component  $b_1(X;\gamma_1)$ but leaving $b_2(W,V)$  unspecified. 
%In this case,  assumption \ref{cndt:cmp} is required for identification, and a consistent and asymptotically normal estimator of $\gamma_1$ can be obtained by semiparametric methods such as  sieve estimation \citep{ai2003efficient}. In the Appendix, we briefly illustrate such an estimator. 

Typically, the dimension of $q$ must be at least  as large as that of $\gamma$.   
For instance, if  $b(W,V,X;\gamma)=(1,X,V^\top,W)\gamma$,   one can use 
$q(X,V,Z)=(1,X,V^\top,Z)^\top$ for the GMM.
\cite{cui2020semiparametric} develop the semiparametric theory for double negative controls by assuming the existence of both the negative control outcome bridge function $b(W,X)$ and negative control exposure bridge function $e(Z,X)$. Their semiparametric efficient estimator is partially based on the above negative control estimator \eqref{eq:gmm} from an unpublished initial draft of the current paper.

The GMM can equally be  applied to   time-series data for parameter estimation \citep[chapter 14]{hamilton1994time}. 
Consider  a typical time-series model,
\begin{eqnarray*}
Y_i= \gamma_0 + \gamma_1 X_i   + U_i   + \varepsilon_{1i},\quad X_i= \alpha_0 +\alpha_1 U_i+  \varepsilon_{2i},\quad U_i=\xi U_{i-1} + (1-\xi^2)^{1/2}\varepsilon_{3i},
\end{eqnarray*} 
with normal white noise $\varepsilon_{1i},\varepsilon_{2i},  \varepsilon_{3i}$. 
As suggested by \cite{flanders2017new}, $Z_i=X_{i+1}$ can be used as a negative control exposure; 
in addition, we use $W_i=Y_{i-1}$  as a negative control outcome, which satisfies
$Z_i\ind (W_i,Y_i)\mid (X_i,U_i)$ and  $W_i\ind X_i\mid U_i$. To estimate $\gamma_1$ via the GMM, we specify a linear  outcome confounding bridge model  $b(W_i,X_i,X_{i-1};\gamma)=(1,X_i,X_{i-1},W_i)\gamma$ and use $q(X_i,X_{i-1},Z_i)=(1,X_i,X_{i-1},Z_i)^\top$ to construct  the   moment restrictions. 
It  seems  surprising that we  can  consistently estimate $\gamma_1$ when    we only observe $X$ and $Y$ but not $U$.
However, this is achieved by selecting appropriate negative control exposure and outcome variables from the observed data  for each observation. 
This approach benefits from the serial correlation of the confounder, but  does not apply to independent observations.

Consistency and asymptotic normality of the GMM estimator have  been established under appropriate conditions in \citet{hansen1982large} and \citet{hall2005generalized}.
Standard errors  and confidence intervals can be constructed based on the normal approximation,
\begin{eqnarray*}
n^{1/2}(\widehat\theta -\theta_0)\stackrel{d}{\rightarrow} N(0,\Sigma_1\Sigma_0\Sigma_1^\top),
\end{eqnarray*}
where $\theta_0$ denotes the true value of $\theta$, and 
\[\Sigma_1= (M^\top\Omega M)^{-1}M^\top\Omega,\quad M=\lim_{n\rightarrow +\infty}\left.\frac{\partial m_n(\theta)}{\partial \theta^\top}\right |_{\theta=\theta_0}, \quad \Sigma_0=\lim_{n\rightarrow +\infty} {\rm Var}\{n^{1/2} m_n(\theta_0)\}.\]
For i.i.d. data, a consistent estimator of the asymptotic variance can be constructed by using
\begin{equation}\label{esti:var}
\begin{aligned}
\widehat \Sigma_1 &= (\widehat M^\top\Omega \widehat M)^{-1}\widehat M^\top\Omega,\quad
\widehat M=\frac{1}{n}\sum_{i=1}^{n}\left.\frac{\partial  h(D_i;  \theta)}{\partial \theta^\top}\right |_{\theta=\widehat \theta},\\
\widehat \Sigma_0 & =\frac{1}{n} \sum_{i=1}^{n} h(D_i;\widehat \theta) h^\top(D_i;\widehat \theta);
\end{aligned}
\end{equation}
and a $95\%$ confidence interval for the elements of   $\theta$ in large samples is  $\widehat \theta \pm 1.96\times \{{\rm diag} (\widehat \Sigma_1\widehat \Sigma_0 \widehat \Sigma_1^\top)/n\}^{1/2}$, where  diag denotes 
the diagonal  elements of  a matrix. 
Variance estimation in the time-series setting is more complicated due to the serial correlation.
When the observed data are serially correlated, $\widehat \Sigma_0$ in \eqref{esti:var} is no longer consistent for $\Sigma_0$, 
and  one should use   heteroscedasticity and autocorrelation covariance (HAC)  estimators that are consistent under relatively weak assumptions \citep{newey1987,andrews1991heteroskedasticity}.
In this paper, we use the Newey-West  estimate of $\Sigma_0$:
\begin{eqnarray*}
\Sigma_0^{\rm HAC}&=& \widehat \Sigma_0 +\sum_{i=1}^{b_n} \{1-\frac{i}{1+b_n}\}(\widehat \Sigma_i + \widehat \Sigma_i^\top), \quad b_n=c\times n^{1/3} {\text{ for some constant c}},\\
\widehat \Sigma_i &=& \frac{1}{n} \sum_{j=i+1}^{ n}  h(D_j;\widehat \theta) h^\top(D_{j-i};\widehat \theta),
\end{eqnarray*} 
where  $b_n$ is  the bandwidth parameter  controlling the number of auto-covariances included in the HAC  estimator; for  practical guidance for the choice of $b_n$, see \cite{andrews1991heteroskedasticity} and \citet[Section 3.5.3]{hall2005generalized}. 
In contrast to the i.i.d. setting, the HAC estimator includes extra covariance terms $\{\widehat \Sigma_i,i\neq 0\} $ to account for the serial correlation.

\section{Positive control outcome}
The negative control outcome assumption, $W\ind X\mid U $, is not met  when the  auxiliary outcome $W$ is causally affected by $X$.  In this case,    we call $W$ a positive control outcome. Let $W(x)$ denote the potential outcome of $W$ when $X$ is set to $x$;   the following assumption  preserves  U-comparability but accommodates a nonzero causal effect of $X$ on $W$, see Figure \ref{fig:M} for a DAG model. 
\begin{assumption}[Positive control outcome]\label{assump:pos}  
$W(x)\ind X\mid U$ for all $x$.
\end{assumption}

\begin{figure}[h]
\centering
\subfloat[Negative control]{
\begin{tikzpicture}[scale=0.75,
->,
shorten >=2pt,
>=stealth,
node distance=1cm,
pil/.style={
	->,
	thick,
	shorten =2pt,}]
%\node [gray](X)  at (3,-3) {$\bm{X}$};
\node (X) at (1.5,0) {$X$};
\node (Y) at (4.5,0) {$Y$};
\node (Z) at (0,2.5) {$Z$};
\node (U) at (3,2.5) {$U$};
\node (W) at (6,2.5) {$W$};
\foreach \from/\to in {X/Y,Z/X,W/Y,U/X,U/Y,U/Z,U/W}
\draw [](\from) -- (\to);  
%\foreach \from/\to in {X/Y1,X/Y2,X/Y3,X/R1,X/R3}
%\draw [->, gray] (\from) -- (\to);  
\end{tikzpicture}
}
\hspace{0.5in}
\subfloat[Positive control]{
\centering
\begin{tikzpicture}[scale=0.75,
->,
shorten >=2pt,
>=stealth,
node distance=1cm,
pil/.style={
	->,
	ultra thick,
	shorten =2pt,}]
\node (X) at (1.5,0) {$X$};
\node (Y) at (4.5,0) {$Y$};
\node (Z) at (0,2.5) {$Z$};
\node (U) at (3,2.5) {$U$};
\node (W) at (6,2.5) {$W$};
\foreach \from/\to in {X/Y,Z/X,W/Y,U/X,U/Y,U/Z,U/W}
\draw [](\from) -- (\to);     
\draw[->, dashed] (X) -- (W);
\end{tikzpicture}
%\label{fig:M}
}
\caption{DAG models   for negative and positive controls. The dashed arrow indicates a possibly nonzero causal effect of $X$ on $W$.} \label{fig:M}
\end{figure}

\begin{proposition}\label{prop:idnp}
Given the latent ignorability Assumption \ref{assump:cnfd}, the outcome confounding bridge Assumption \ref{assump:bridge}, and the positive control Assumption \ref{assump:pos},  then $E\{Y(x)\} =  E \{b(W(x),x)\}$ for all $x$.
\end{proposition}
The potential outcome mean $E\{Y(x)\}$   depends on the distribution of $W(x)$ rather than the observed distribution of $W$.
Given a positive control outcome and a negative control exposure, \eqref{eq:intg} still holds, and thus can be used to identify the outcome confounding bridge.
As a consequence, the causal effect of $X$ on $Y$  can be identified if  both a positive control outcome and a negative control exposure are available and  the causal effect of $X$ on $W$ is known a priori. 
Suppose the bridge function has an additive form $b(W(X),X;\gamma)= b_1(X;\gamma_1) +b_2(W(X))$ 
where  the structural parameter  $\gamma_1$ is unknown. Then, the potential outcome mean can be rewritten as $E\{Y(x)\} = E\{b_1(X;\gamma_1) + b_2(W(X))\}$. We let $\gamma_2(x) = E\{b_2(W(x))\}$ be a specified functional form
of $x$ in a sensitivity analysis measuring the the mean of $W(x)$ transformed by some function $b_2$.
The estimation is analogous to the GMM method in Section 4. Define the vector of  moment restrictions 
\begin{eqnarray}\label{eq:sensgmm}
h(X_i, Y_i; \gamma_1, \gamma_2(x), \Delta)= \left\{\begin{array}{l}
[Y_i - b_1(X_i;\gamma_1) - \gamma_2(X_i)]\times q(X_i), \\
\Delta -[b_1(x_1;\gamma_1)-b_1(x_0;\gamma_1) + \gamma_2(x_1) - \gamma_2(x_0)],
\end{array}\right.
\end{eqnarray}
with a user-specified vector function $q$. 
The first component in \eqref{eq:sensgmm} consists of unbiased  estimating equations for $\gamma_1$ because
$E\{Y-b_1(X) - \gamma_2(X)\mid X\}=0$, and the second one for $\Delta$.
In practice, the users can make use of auxiliary information of $\gamma_2(x)$ if possible or specify a functional form based on expert knowledge to test the robustness of the estimation method  against the effect size on the positive control.
We further illustrate this with the following examples.

\begin{example}\label{exmp:pos}
Consider  binary exposures $(X,Z)$ and the linear outcome confounding bridge $b(W(X),X)=\gamma_0+\gamma_1X +\gamma_2 W(X)$ for a positive control outcome $W$. 
Then  $E\{Y(x)\} = \gamma_0+\gamma_1x  + \gamma_2E\{W(x)\}$ and
${\rm ACE}_{XY}= \gamma_1 + \gamma_2 \times {\rm ACE}_{XW}$.
Identification of $(\gamma_1,\gamma_2)$ is identical as in the negative control outcome case, with
$ \gamma_2=E({\rm RD}_{ZY\mid X})/E({\rm RD}_{ZW\mid X})$ and $\gamma_1= E({\rm RD}_{XY\mid Z}) - \gamma_2\times E({\rm RD}_{XW\mid Z})$.
In contrast with the negative control setting in Example \ref{exmp:neg2}, identification with a positive control outcome involves the average causal effect of $X$ on $W$.
Using ${\rm ACE}_{XW}$ as a sensitivity parameter,  sensitivity analysis can be performed to evaluate the plausibility of a  causal effect of  $X$ on $Y$; 
if ${\rm ACE}_{XW}$ is  known to belong to the interval $[a,b]$, then the bound for ${\rm ACE}_{XY}$ is  $[\gamma_1+\gamma_2 a,\gamma_1+\gamma_2 b]$;  
given the sign of $\gamma_2$, the sign of $E({\rm RD}_{XY\mid Z}) - {\rm ACE}_{XY}$, i.e.,  the confounding bias,  can be inferred from the sign of   $E({\rm RD}_{XW\mid Z}) - {\rm ACE}_{XW}$.
\end{example}

\begin{example}\label{exmp:smk}
In studies assessing the effect of intrauterine smoking ($X$) on offspring  birthweight ($Y$) and seven years old body mass index  ($W$),  \cite{dsmith2008assessing,dsmith2012} used paternal smoking  ($Z$) as a negative control exposure, and observed that  
\[
\begin{aligned}
&E({\rm RD}_{XY\mid Z}) =-150 \text{ g},  \quad && E({\rm RD}_{XW\mid Z})=0.15 \text{ kg/m$^2$},\\  
&E({\rm  RD}_{ZY\mid X})=-10 \text{ g},   \quad && E({\rm RD}_{ZW\mid X}) =0.11 \text{ kg/m$^2$}.   
\end{aligned}
\]
Following the analysis in Example \ref{exmp:pos},
we obtain $\gamma_2=-91,\gamma_1=-136$, and thus
${\rm ACE}_{XY} = -136-91\times {\rm ACE}_{XW}  \text{ g}$.
A necessary condition to explain away the observed  impact of   intrauterine smoking  on birthweight (i.e., to make ${\rm ACE}_{XY} \geq 0$) is  ${\rm ACE}_{XW} \leq -1.5$ kg/m$^2$,  a protective  effect of  intrauterine smoking on later-life body mass index.
However, intrauterine smoking  is unlikely to have such a considerable protective effect against obesity, and in fact, researchers have hypothesized although not definitely established that intrauterine smoking is likely to increase not decrease the risk of offspring obesity \citep{al2006does}.
Therefore,  the most plausible explanation is that  intrauterine smoking  decreases  offspring birthweight,  at least $-136$g on average if one believes intrauterine smoking  can also cause offspring adiposity.

\end{example}

\section{Simulation studies}

\subsection{Simulations for a binary exposure}

We provide two simulation examples in this and the next section.
In the first simulation, we generate two variables $V,U\thicksim N(0,1)$ with correlation $\sigma_{uv} =0.5$. Then we generate the negative control exposure, negative control outcome based on the following models $Z=0.5+0.5V+U + \varepsilon_1, W=1-V+\xi U+\varepsilon_2$ with $\varepsilon_1,\varepsilon_2\thicksim N(0,1)$. The exposure and the potential outcome are generated based on 
$\logit\{\pr(X=1\mid Z,V,U)\}=-0.5+Z+0.5V+ \eta U, Y(x)=1+0.5 x+2V+U+1.5xU + 2\varepsilon_2$ with $\eta$ encoding the magnitude of confounding and $\xi$  the association between the negative control outcome  and the confounder.
We analyze  data with the negative control approach (NC), standard  inverse probability weighting (IPW), and   ordinary least square (OLS).

\begin{figure}[ht]
\includegraphics[scale=0.3]{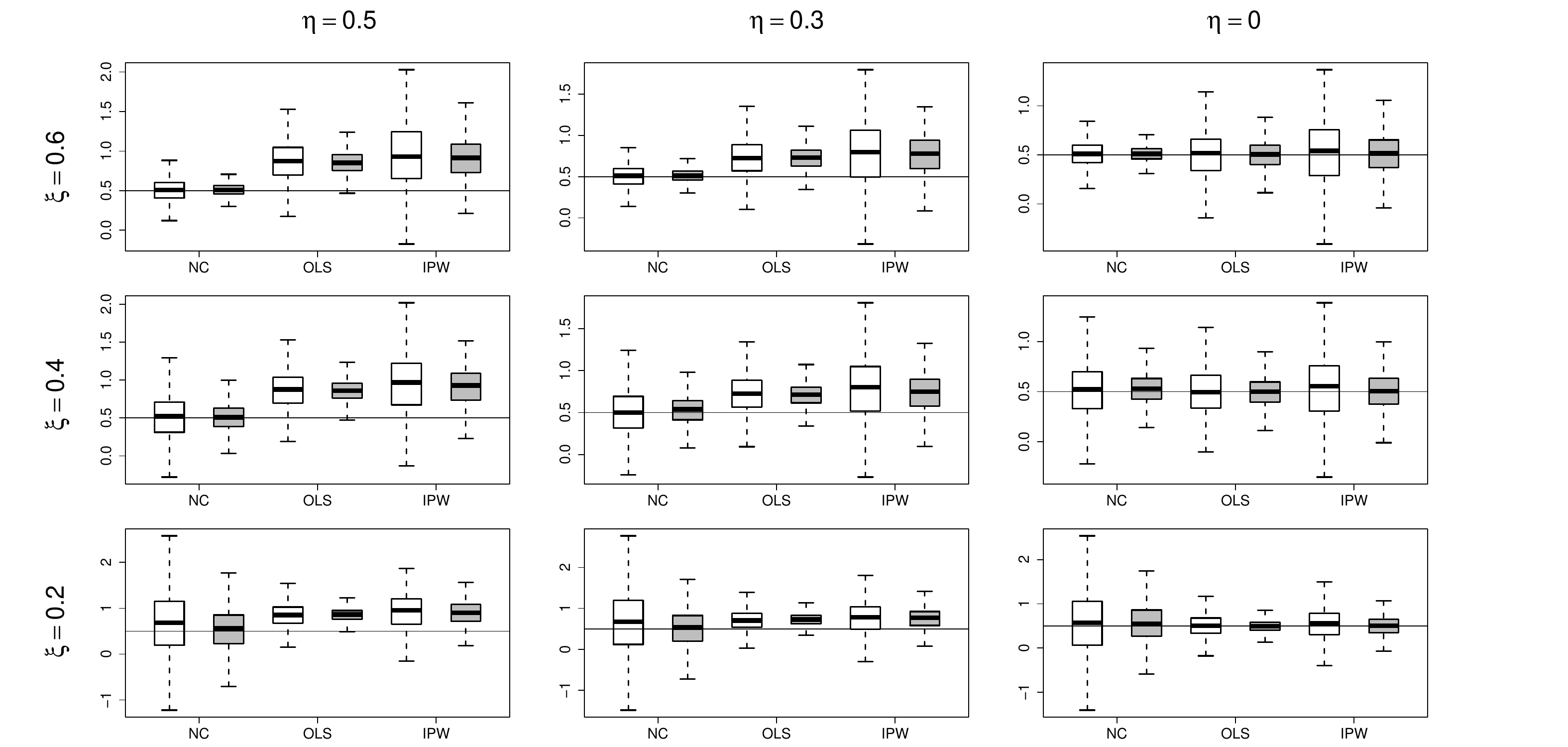}
\caption{Boxplots for estimators of the average causal effect.} \label{fig:ace}
{\footnotesize\begin{flushleft}
Note: For NC, $b=(1,X,V,W,XV,XW)\gamma$ and $q=(1,X,V,Z,XV,XZ)^\top$ are used for the GMM; 
for IPW, a logistic model for $\pr(X=1\mid V)$ is used; for OLS, a linear  outcome model is used.
White boxes are for sample size $500$ and gray ones  $1500$; the horizontal line  marks   the true value of the average causal effect. 
\end{flushleft}}
\end{figure}

For each choice of $\eta=0,0.3,0.5$ and $\xi=0.2,0.4,0.6$, we replicate  $1000$ simulations  at sample size $500$ and $1500$, respectively, and  summarize  results  as boxplots  in Figure \ref{fig:ace}.
From Figure \ref{fig:ace}, the negative control estimator has small bias in all  settings; in contrast,   ordinary least square   and  inverse probability weighted estimators are biased except under no unmeasured confounding ($\eta=0$).
When the association between the negative control outcome and the confounder is moderate to strong ($\xi=0.4,0.6$), the  negative control estimator  is more efficient than the other two, but  has greater variability otherwise ($\xi=0.2$).
Table \ref{tbl:ace} presents  coverage probabilities  of  $95\%$ negative control confidence intervals based on a normal approximation, which  generally   approximate the nominal level of $0.95$.
But,  when the association between the  negative control outcome and the  confounder is weak ($\xi=0.2$), the coverage probabilities are slightly inflated.
Therefore, we recommend the negative control approach to  remove the confounding bias in observational studies, and to enhance efficiency, 
we recommend when possible  to use a negative control outcome that is strongly associated with the confounder. 

\begin{table}
%\centering
\tbl{Coverage probability of $95\%$ negative control confidence interval for the average causal effect.}
{
\begin{tabular}{rcccccccccccc}
\toprule
&& &\multicolumn{2}{c}{$\eta=0.5$}& &\multicolumn{2}{c}{$0.3$}&& \multicolumn{2}{c}{$0$}\\
\cline{4-5}\cline{7-8} \cline{10-11}
&$n=$&      &500&1500&&	500&1500&&	500&	1500\\
\midrule
\multirow{3}{1em}{$\xi=$} &0.6&      &0.945&0.936&&	0.958&0.953&&	0.954&	0.935\\
         &$0.4$     &   &0.958&0.957&&	0.968&0.955&&	0.964&0.956\\
         &$0.2$        &&0.953&0.963&&	0.970&0.963&&	0.978&0.979\\
\bottomrule
\end{tabular}}
\label{tbl:ace}
\end{table}

\subsection{Simulations for time series data}
We generate   time-dependent data according to  
\begin{eqnarray*}
&& U_i=\xi U_{i-1} + (1-\xi^2)^{1/2}\varepsilon_{1i},\quad V_i=0.6U_i +  \varepsilon_{2i},\quad X_i =0.4 + 1.5V_i + \eta U_i +\varepsilon_{3i},\\
&&Y_i= 0.5 + 0.7X_i + 1.5V_i+0.9U_i   +\varepsilon_{4i},\quad \varepsilon_{1i},\varepsilon_{2i},  \varepsilon_{3i},\varepsilon_{4i}\thicksim N(0,1),
\end{eqnarray*}
where $U_i$ is  a stationary autoregressive  process with autocorrelation coefficient $\xi$, and $\eta$ controls the magnitude of confounding.
We analyze  data with the negative control approach (NC),   ordinary least square (OLS) without controlling lagged exposures, and lagged-OLS by  controlling   one-day lagged  exposure. For  the negative control approach, we use $W_i=Y_{i-1}$ and $Z_i=X_{i+1}$  as  negative controls, and  do not need auxiliary data.

\begin{figure}[ht]
\includegraphics[scale=0.3]{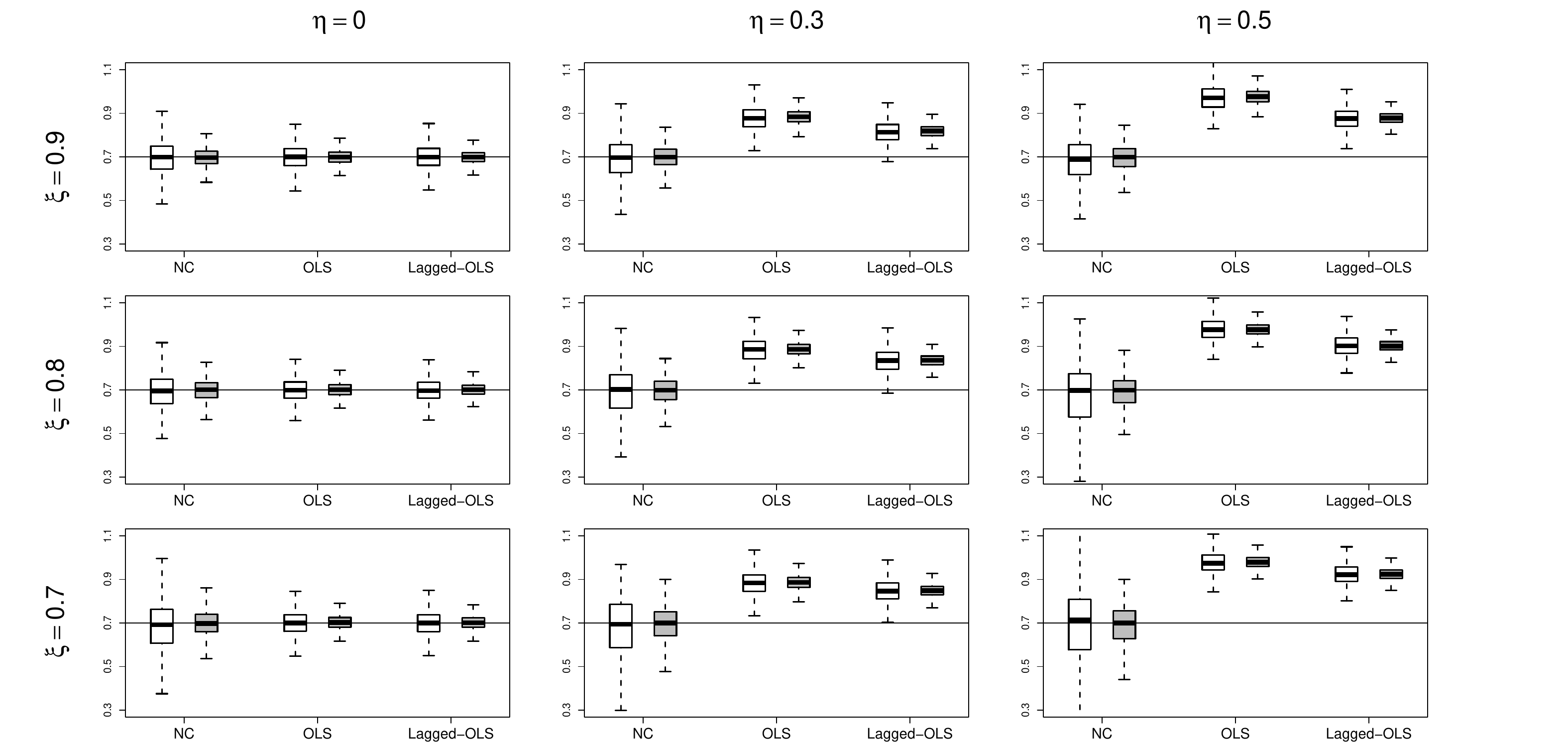}
\caption{Boxplots for time series data analysis.} \label{fig:time}
{\footnotesize\begin{flushleft}
Note: For NC,  $b= (1,X_i,X_{i-1},V_i,V_{i-1},W_i)\gamma$ and  $q=(1,X_i,X_{i-1},V_i,V_{i-1},Z_i)^\top$
are used  for the GMM.
White boxes are for sample size $500$ and gray ones  $1500$; the horizontal line  marks   the true value of the structural parameter.
\end{flushleft}}

\end{figure}

For each choice  of $\xi=0.7,0.8,0.9$ and $\eta=0,0.3,0.5$, we  replicate  $1000$ simulations  at sample size $500$ and $1500$, respectively.
Figure \ref{fig:time} presents   boxplots of  the estimators.
The negative control estimator has small bias in all  nine scenarios, and its variability becomes smaller  as  autocorrelation  of the confounder process increases. 
The $95\%$  negative control confidence intervals based on the \cite{newey1987} variance estimator   have    coverage probability approximating  $0.95$, 
as shown in Table \ref{tbl:time}.
The ordinary least square estimator is biased except under no unmeasured confounding ($\eta=0$), in which case, it is more efficient than the negative control estimator.
Controlling lagged exposures in  ordinary least square  can reduce  confounding bias, but cannot eliminate it.
Therefore,  we recommend the negative control approach for estimation of a linear time-series regression model in the presence unmeasured confounding.

\begin{table}[ht]
\tbl{Coverage probability of $95\%$ negative control confidence interval for  the time-series model.}
{
\begin{tabular}{rlccccccccccc}
\toprule
&& &\multicolumn{2}{c}{$\eta=0$}& &\multicolumn{2}{c}{$0.3$}&& \multicolumn{2}{c}{$0.5$}\\
\cline{4-5}\cline{7-8} \cline{10-11}
&$n=$&      &500&1500&&	500&1500&&	500&	1500\\
\midrule
\multirow{3}{1em}{$\xi=$}
&$0.9$&&0.953& 0.947&&	0.948&	0.950&&	0.950&	0.947  \\
&$0.8$&&0.979&	0.952&&	0.952&	0.943&&	0.933&  0.946 \\
&$0.7$&&0.982&	0.974&&	0.937&	0.942&&	0.912&   0.940\\
\bottomrule
\end{tabular}}
\label{tbl:time}
\end{table}

\section{Evaluation of the effect of air pollution  on mortality }
While there are many long-term threats posed by air pollution, its acute effects on mortality also pose an important public health concern. 
We apply the negative control approach  to evaluate the short-term effect of air pollution on mortality using  datasets  from  a time-series  study
in Philadelphia, New York, and Boston.
Here we present the analysis results for Philadelphia and relegate those for the other two cities to the Supplementary Material.
The dataset for Philadelphia contains $n=2621$ daily records of PM2.5, temperature, ozone, date, and   number of deaths in Philadelphia  from 1999  to 2006.
With accidental deaths excluded, the number of deaths  ranges from  73 to 179, which is  often assumed to have a Poisson distribution.
In our analysis, we use square root of the number of deaths  for the purpose  of normalization and variance stabilization  \citep{freeman1950transformations}. 

For a given day $i$, we let $Y_i$ denote the square root of  number of deaths, $X_i$  be the PM2.5 concentration measurement,
$V_i$ consist of  temperature and its square,   ozone, and $X_{i-1}$ to control lagged effects, and $T_i$ consist of   polynomial and Fourier   bases  of time to account for both   secular and seasonal trends: 
\[T_i=\{i/n, i^2/n^2,\sin(2\uppi i/365),\cos(2\uppi i/365),\ldots,\sin(8\uppi i/365),\cos(8\uppi i/365)\}.\] 
We assume a linear outcome model, $Y_i=\beta_1 X_i + (1,V_i,T_i) \beta_2 + U_i$,
and we are interested in the regression coefficient $\beta_1$ that encodes the immediate effect of current day PM2.5 on mortality.
All results are summarized in Table \ref{tbl:phila}, where confidence intervals and p-values  are obtained  from  the  normal approximation  and  the \cite{newey1987}  variance estimator  is used   to account for serial correlation.
A standard regression analysis shows that  short-term exposure to  PM2.5 can  significantly increase mortality,  with  point estimate $0.0084 $ and  $95\%$ confidence interval  $(0.0048,  0.0120)$ for $\beta_1$.  
%Using  $Z_i=X_{i+1}$ as an instrumental variable leads to  a larger estimate of $\beta_1$   but  a much wider   confidence interval.
However,  a confounding test by fitting  the model 
\[W_i=\alpha_1 X_i+\alpha_2 Z_i+ (1, X_{i-1},V_{i-1},T_{i-1})  \alpha_3 + U_{i-1},\]
with $W_i=Y_{i-1}$,  results in  point estimate  $-0.0040$ of $\alpha_1$ with $95\%$ confidence interval   $(-0.0073, -0.0007)$ and p-value $0.0167$, 
and point estimate $0.0041$ of $\alpha_2$ with $95\%$ confidence interval  $(0.0011,  0.0071)$    and p-value $0.0072$.
These results  suggest presence of  unmeasured confounding  because  $W_i$ occurs before $X_i$ and $Z_i$, and should not be  affected by them.
%In this case,  instrumental variable estimation is also problematic, because $Z_i$ is   confounded due to an association  with $U_i$ through $U_{i+1}$. 
Thus, ordinary least square appears not entirely appropriate in this setting.
We apply the proposed negative control approach and use $Z_i=X_{i+1}$ and $W_i=Y_{i-1}$ as the  negative control exposure and outcome, respectively.
We assume a linear   outcome confounding bridge 
$b=(1,X_i,V_i,V_{i-1},T_i,W_i)\beta$, and use $q=(1,X_i,V_i,V_{i-1},T_i,Z_i)^\top$ for the GMM.
Compared to the standard regression, the negative control estimate  of $\beta_1$ is  attenuated toward zero a lot, although 
it  still  has some significance  with  point estimate $0.0045$ and    $95\%$ confidence interval $(-0.0006,  0.0097)$.
Further  analyses  controlling   longer lagged  exposures  by including  $X_{i-2}$ and $X_{i-3}$ in $V_i$ lead to analogous results as those obtained 
when only $X_{i-1}$ is controlled. Our analyses indicate  presence of  unmeasured confounding in the air pollution study in Philadelphia.
In parallel analyses we provide in the Supplemental Materials, unmeasured confounding is also detected in  the dataset for New York via the negative control approach,
but not  detected in the dataset for Boston.
After accounting for unmeasured confounding, our negative control inference shows a significant acute effect of PM2.5 on mortality in Philadelphia,
but such an effect is not detected in  New York or Boston.

\begin{table}
\tbl{Estimates and 95\% confidence intervals ($\times 10000$) of the effect of air pollution  in Philadelphia.}
{
\begin{tabular}{lllcllcllc}
\toprule
&&\multicolumn{8}{c}{Number of lagged exposures controlled}\\
&&\multicolumn{2}{c}{One day }&&\multicolumn{2}{c}{Two days }&&\multicolumn{2}{c}{Three days }\\
\toprule
&  & Estimate & $p$-value&   & Estimate &  $p$-value& &Estimate&  $p$-value\\
\midrule
\multicolumn{4}{l}{Ordinary least square}\\
$\beta_1$&& 84 (48,  120) &  0     && 78 (41,  115) &  0     && 79 (43,  116) &  0\\[12pt]
%\midrule
%\multicolumn{4}{l}{Instrumental variable estimation}\\
%$\beta_1$&& 45 (-27,  118) &  0.2213&& 41 (-32,  114) &  0.2712&& 40 (-33,  113) &  0.2829\\
%\midrule
\multicolumn{4}{l}{Confounding test}\\
$\alpha_1$&&-40 (-73, -7)   &  0.0167&&-39 (-71, -7)  &  0.0174&&-40 (-72, -7)   &  0.0158\\
$\alpha_2$&& 41 (11,  71)  &  0.0072&& 40 (10,  69)  &  0.0080&& 39 (10,  69)  &  0.0083\\[12pt]
%\midrule
\multicolumn{4}{l}{Negative control estimation}\\
$\beta_1$&& 45 (-6,  97)   &  0.0854&& 46 (-6,  98)  &  0.0844&& 46 (-7,  99)  &  0.0915\\
\bottomrule
\end{tabular}}
\label{tbl:phila}
\end{table}

\section{Discussion}
We propose an outcome confounding bridge  approach for    negative control/proximal   inference on causal effects.
We clarify the key assumptions  and the roles of negative control outcome and exposure, and discuss   robustness  and sensitivity  of the approach.
% Our approach enjoys the ease of implementation of standard parametric inference methods such as the GMM and two stage least square.
%Sometimes, it is of interest  to consider a semiparametric or nonparametric outcome confounding bridge, in which case,  semiparametric methods such as  sieve estimation \citep{ai2003efficient} can be applied.
In the supplementary material, we  provide some insights on the   connection between    the negative control  and the  instrumental variable   approaches, focusing on   estimation of  a structural model.
As we illustrate, an invalid instrumental variable that fails to be independent of the unmeasured confounder can be viewed as a negative control exposure,
and   a  negative control outcome can be used to repair such an invalid IV by applying our double negative control adjustment.
Under a linear structural model, we show  double robustness property of the negative control estimator, in the sense that it is consistent if either the confounding bridge is correctly specified or the negative control exposure is a valid IV.

Besides for causal effect evaluation, our approach has  important implications for the design of observational studies. Even if an exposure or response factor is not directly relevant to the study variables in view,  it is useful to collect them and use them as negative controls for the purpose of  confounding diagnostic and  bias adjustment. 
Time-series studies, such as the air pollution example we consider,  are particularly well-suited for the proposed negative control approach,
because   negative controls can be constructed from observations of  the exposure and outcome themselves.
However, in general, our approach requires one to collect extra data about   negative control variables.
%For the instrumental variable design, we recommend that one  collects  negative control outcomes to enhance  robustness of  IV estimation.

The negative control assumptions we present in this paper describe the general principles for selecting   negative control variables, 
and the examples we give provide  guidance for certain specific studies;
but in general,  subject matter knowledge about the data generating mechanism and  the potentially unmeasured confounders,  
such as specificity of  the exposure-outcome relation \citep{hill1965environment,lipsitch2010negative}, is indispensable to choose an appropriate negative control.

Our approach has promising application in modern big and multi-source data analyses. 
Identification of the outcome confounding bridge and the average causal effect depends only on 
$\pr(Y,Z,X)$ and $\pr(W,Z,X)$ but not the joint distribution of $(Y,W)$,  and thus enjoys the convenience of  data integration and two-sample inference.
For certain outcome confounding bridge  models such as the linear  one, estimation  of the average causal effect  requires only summary  but not individual-level data,
and thus allows for synthetic analysis  by using results  from multiple studies. Such extensions will be carefully developed in the future.

\section*{Supplementary material}
Supplementary material online  includes discussion about the connection between    the negative control  and the  instrumental variable,   additional simulation with a continuous exposure, proofs of propositions   and  theorems, details for examples,
and analysis results for the effect of air pollution in New York and Boston.

%\section*{Acknowledgments}
%We are grateful for valuable comments from the editor and two anonymous reviewers. This work was partially supported by National Key R\&D Program of China (2020YFE0204200, 2022YFA1008100), and National Natural Science Foundation of China (12071015, 12292983).

\bibliographystyle{apalike}

\bibliography{CausalMissing}

\end{document}